# POLYMORPHIC TRANSITION IN P-DIHLORBENZOL NANOPARTICLES

**M.A.Korshunov**
***Institute of physics it. L. V.Kirenskogo of the Siberian separating of the Russian Academy of Sciences, 660036 Krasnoyarsk, Russia***
*e-mail: mkor@iph.krasn.ru*

**Abstract.** We have obtained experimentally the low frequency Raman spectra of p-dihlorbenzol nanoparticles. Nanoparticle sizes are determined with the help of electron microscope. It was found that in the lattice vibration spectra from 70 nm the summary spectrum of α-paradihlorbenzola and β-paradihlorbenzol structures appears. It agrees with both calculations of nanoparticle structure by molecular dynamics and calculations of spectra histograms of the lattice vibrations by a Dyne's method.

In the molecular crystals, the transition from massive particles to nanoparticles is accompanied by change of continuances of a lattice and orientation of molecules [1]. It can affect change of structure of a lattice and lead to phase transition. These phenomena should find reflection in change of dynamics of a lattice and accordingly in spectrums of the lattice oscillations. For studying of this question in the present operation comparison of the observational spectrums of a Raman effect of light of the lattice oscillations of nanoparticles of a p-dihlorbenzol (which has depending on temperature three updatings α, β and γ [2]) with settlement spectrums of nanoparticles has been spent at change of their size. For interpretation of a spectrum of the lattice oscillations calculations on a method the Dyne [3] are carried out. This method allows to calculate the spectrums of disorder structures and gives the chance to judge lattice parametres, dynamics of molecules and their arrangement. The nanoparticle structure was modelled by means of a method of the molecular dynamics [4]. As object of examination the organic molecular crystal of paradihlorbenzoles as it combines all necessary properties which are supposed to be explored has been chosen. Massive monocrystals of a p-dihlorbenzol are well enough studied by various methods. There is X-ray diffraction data [5] and interpretation of the lattice oscillations of this monocrystal [6].

Nanoparticles of a p-dihlorbenzol of the necessary size have been obtained. The sizes of nanoparticles were spotted on an electronic microscope. Record of spectrums of a Raman effect of light of nanoparticles was done on spectrometer Jobin Yvon T64000. The obtained spectrums of a Raman effect of light of small frequencies of nanoparticles 70 nm are presented in Figures 1, and values of frequencies (ν) the most intensive lines of a spectrum in the table. At reduction of the sizes of particles in a spectrum occurrence of lines caused by a circumrotatory spectrum of a Raman effect of molecules of air that was considered in the given spectrums in Figure 1 is observed. These lines have been subtracted from an initiating spectrum.

Using a method of the molecular dynamics and calculation of spectrums of the lattice oscillations it is possible to study structure and processes of formation of a nanoparticle at level of a motion of separate molecules. The method of the molecular dynamics allows to build trajectories of particles, using the equations of a classical mechanics [4]. The High light in this method is calculation of forces.

In the molecular structures unlike the structures consisting of atoms, the potential energy depends not only on a relative positioning of centre of gravity of molecules, but also from orientation of each of molecules that considerably increments time of calculations. The structure of molecules was accepted by terrain clearancely rigid. The interaction potential has been chosen in shape (6-EXP) [7]. In calculations coefficients in an interaction potential, obtained by us [8] were used earlier. For calculation of velocities and co-ordinates of molecules on a method of the molecular dynamics the algorithm of Verle in the velocity shape [9] was used. Thus quantity of velocity for different directions set proceeding from a task in view.

As it is known the potential of a crystal depending on orientation of molecules and their arrangement can have some local minimums, but to the true structure there corresponds least of them. It can be used for a finding of structure of nanoparticles with use of a method of the molecular dynamics. At first any structure of a particle is set. Then, changing Euler's spotting orientations of molecules and changing an arrangement of centre of gravity of molecules the corners, we find a curve of potentials of local minimums in a crude approximation. The method of the molecular dynamics is applied to structure answering to least of minimums allowing to improve a motion of molecules and to spot structure more precisely. The structure of an one-layer molecular film of a p-dihlorbenzol at first paid off at a small amount of molecules with magnification of quantity of stratums this film forms a nanoparticle.

Using the found structure, calculations of spectrums of the lattice oscillations on a method the Dyne have been carried out.

Idea of computing procedure on this method the following. Being set by an arrangement of atoms in a molecule and considering fields of intermolecular forces, we gain devices of a dynamic matrix. Power stationary values were by direct differentiation of the potential energy which has been written down in the form of steam rooms atom-nuclear of interactions. The matrix order is equal to number of molecules accepted in viewing increased on six. For calculation of a spectrum of disorder structures good effects are given by a method the Dyne. On the basis of calculations histograms which show probability of display of lines of a spectrum in the chosen frequency interval have been obtained. In more details the method is featured in operation [3]. The calculated histograms of spectrums are presented in Figure 2.

The p-dihlorbenzol monocrystal at a room temperature crystallises in space group $P2_1/a$ with two molecules in an unit cell [5]. In a spectrum of the lattice oscillations of a monocrystal it should be observed six intensive lines of the molecules caused by circumrotatory huntings round axes of main moments of inertia. At temperature 303.9T there is a polymorphic transition in β - a phase with one molecule in an unit cell and space group P1. In a spectrum of the lattice oscillations of a monocrystal it should be observed three intensive lines related to circumrotatory huntings of molecules of a crystalline lattice. Is available as γ - a phase at temperature more low 273T with two molecules in an unit cell and a space group $P2_1$/with [2].

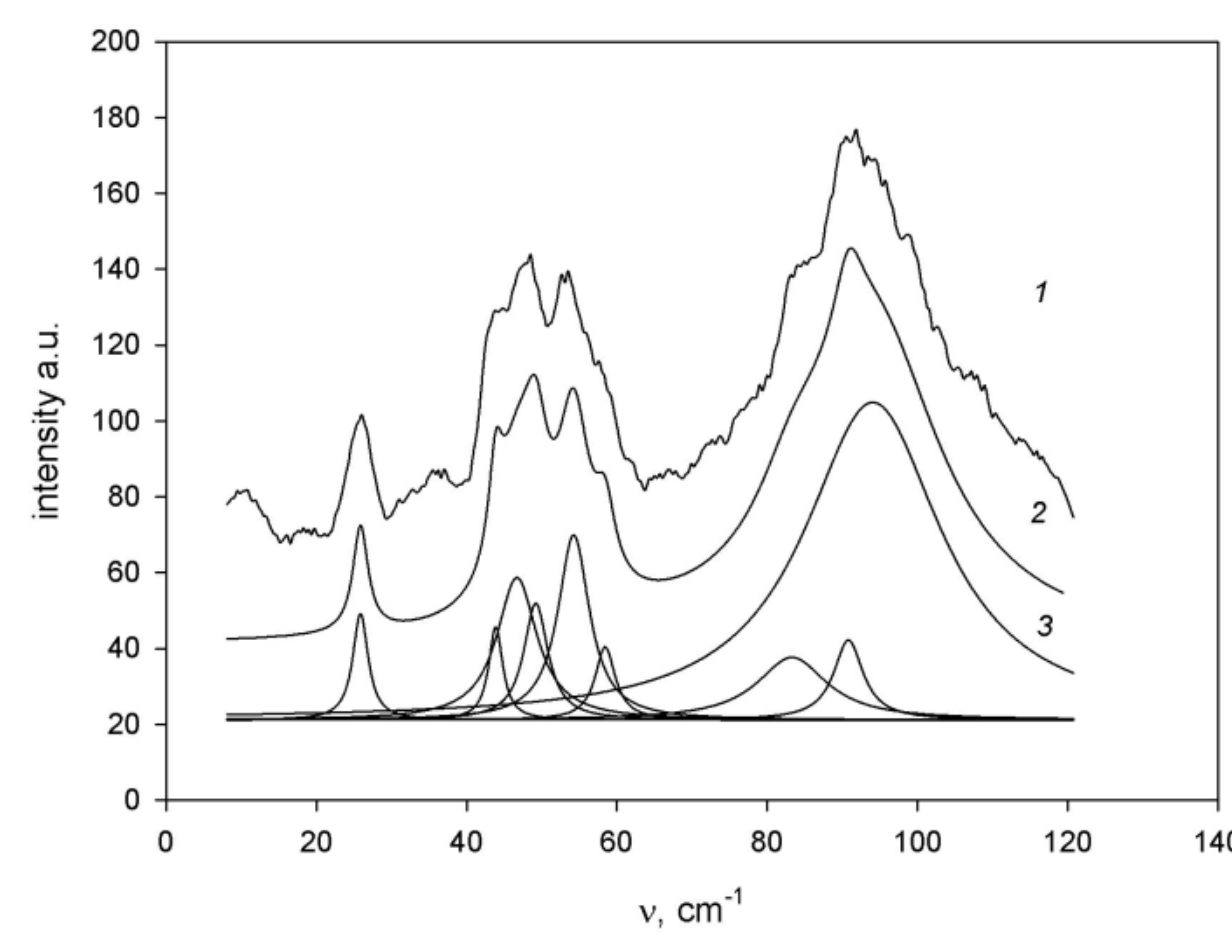

**Figure 1.** The observational spectrum of the lattice oscillations of a p-dihlorbenzol at the size of particles of 70 nanometers (*1*), effect of partitioning of lines (*3*) observational spectrums (*1*) and their net spectrum (*2*).

In Figure 1 the spectrums of the lattice oscillations obtained for particles in the size of 70 nanometers of a p-dihlorbenzol. Apparently in a spectrum it is observed more than six lines unlike spectrums of massive samples. Besides in spectrums a series of lines of small intensity is scored also. In the table values of frequencies of intensive lines of a p-dihlorbenzol (ν) are given at the size of particles of 70 nanometers. In Figures the effect of partitioning of lines (*3*) observational spectrums (*1*) and their net spectrum (*2*) also is given. It is visible that the net spectrum is close to the observational spectrum. Apparently from Figure 1 in a spectrum it is observed nine intensive lines. Six from which belong to a spectrum of the lattice oscillations α - p-dihlorbenzol phases (values of frequencies of the six first lines in the table) and three lines correspond to frequencies β - p-dihlorbenzol phases (last three values of frequencies in the table) [6,8]. At comparison of value of frequencies of respective lines in the massive sample and in a spectrum of nanoparticles of 70 nanometers they decrease. Spectrum change at reduction of the sizes of nanoparticles can be related to reorganisation of structure of a lattice and change of its parametres. Occurrence is probably related to it in spectrums of additional lines of small intensity of the flaws of a lattice caused by presence in particular vacancies also.

**Table.** Frequencies of lines of a p-dihlorbenzol ν ($cm^{-1}$) at the size of nanoparticles 70nm.

| Frequency (ν) |
|---|
| 26.0 |
| 46.0 |
| 48.5 |
| 52.5 |
| 91.0 |
| 97.5 |
| 43.5 |
| 55.0 |
| 83.5 |

Calculations of an one-layer molecular film of a p-dihlorbenzol (30×30 molecules) have shown that molecules settle down in the core parallelly planes of benzene rings to each other that is this structure is close to structure β - p-dihlorbenzol phases. At addition of a following stratum cross orientation of molecules changes a little. At magnification of stratums more than 30 in volume образовывается the ranked crystalline structure similar to a monocrystal α-paradihlorbenzola but is maintained by places structure β - phases. On boundaries of perpendicular section of a nanoparticle of a molecule have some orientation and transmitting disorder (~3 stratums of molecules), and at section centre generally structure ranked α-paradihlorbenzola. The short-range order is thus maintained, but there can be a long-range. In a spectrum of a similar particle should be observed not only oscillations from a volume part, but also and the superficial oscillations.

Using the same coefficients in an interaction potential, and the calculated structure of nanoparticles taking into account the orientation disorder and vacancies the histogram on a method the Dyne (Figure 2 (*1*)) has been calculated. As we see, in the core the calculated spectrum is similar to the observational spectrum in Figure 1 (*1*).

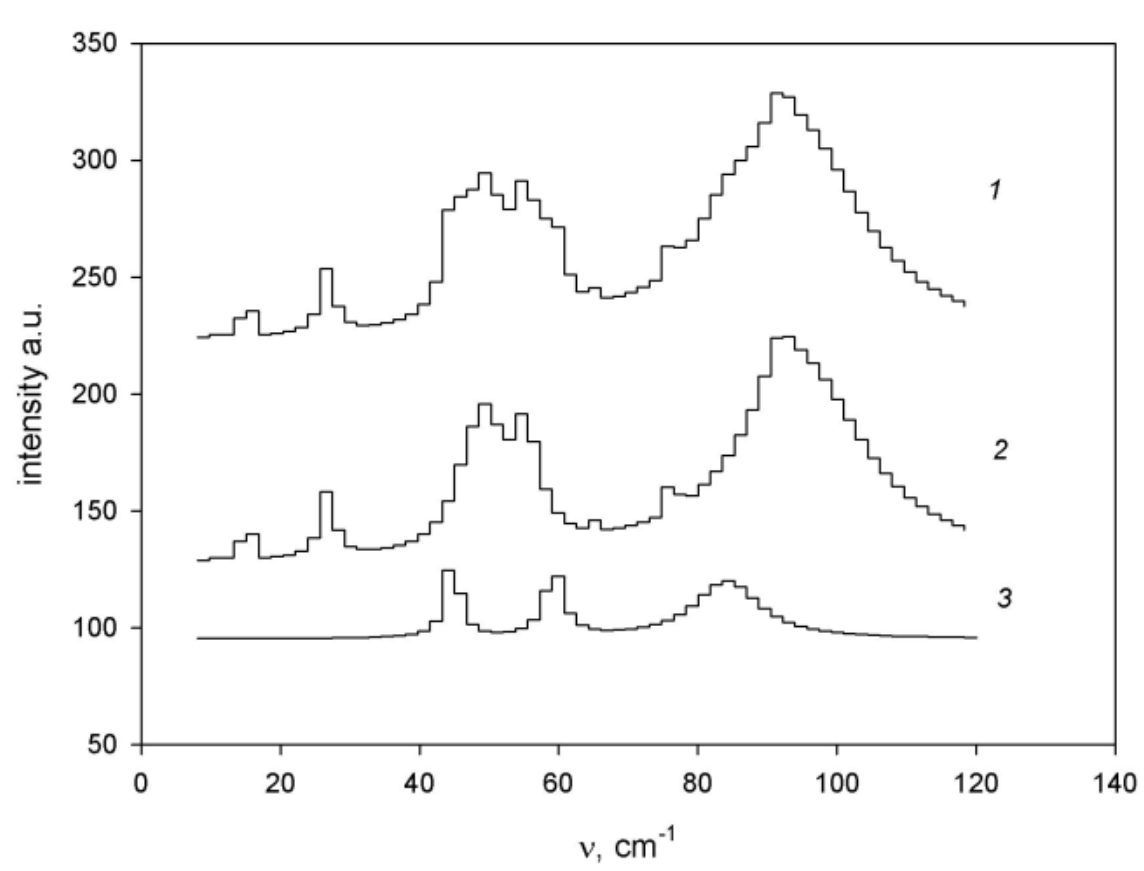

**Figure 2**. Histograms of frequencies of the lattice oscillations of nanoparticles with a size ~ 70 nanometers with structure α and β updatings of a p-dihlorbenzol (*1*), α - a paradihlorbenzol with orientation disorder of molecules and in the presence of vacancies in structure and the account of the superficial oscillations (*2*), nanoparticles with structure β - a paradihlorbenzol (*3*).

Effects of calculation of small clusters have shown that six intensive lines in a spectrum of nanoparticles a paradihlorbenzol (table) are related to orientation oscillations of molecules. Straggling and occurrence of additional lines it is caused by display in a spectrum of presence of the orientation disorder of molecules, vacancies and the superficial oscillations. The account of the superficial oscillations and the orientation disorder of molecules causes straggling of frequencies and occurrence in a spectrum of lines in field below 20.0 $cm^{-1}$. Occurrence of additional lines in the field of 70 $cm^{-1}$ is caused by presence of vacancies. As have shown the calculations, the six first lines in the table are caused α - structure of a nanoparticle of a p-dihlorbenzol calculation of the histogram to which it is presented in Figure 2 (*2*). Three last lines in the table are related to presence β - the phases which effects of calculation are shown in Figure 2 (*3*). Presence in structure of both updatings also causes an observable spectrum in nanoparticles of a p-dihlorbenzol Figure 1 (*1*).

At calculations it is found that there is a magnification of parametres of a lattice of molecules of a p-dihlorbenzol. Reduction of values of frequencies of lines of a spectrum is related to it at reduction of the sizes of nanoparticles. And, apparently, possible phase transition in some fields of a nanoparticle from α - updatings in β - updating, as is observed in the observational spectrums.

The observational spectrums of a Raman effect of light of small frequencies of nanoparticles of a p-dihlorbenzol at the size of 70 nanometers have shown that values of frequencies of lines go down in comparison with massive samples and there is a series of additional lines. At modelling of structure of particles on a method of the molecular dynamics and calculation of histograms of spectrums of the lattice oscillations on a method the Dyne, it is found that at reduction of the sizes of particles lattice parametres are incremented, at nanoparticle structure there are two updatings of a p-dihlorbenzol α and β that finds the reflexion in the observational spectrums. It speaks about that that regarding structure polymorphic transition caused by size effects is observed.